\documentclass[aps, prb, superscriptaddress, twocolumn]{revtex4-2}
\usepackage[utf8]{inputenc}

\usepackage{amsmath}
\usepackage{hyperref}
\usepackage{graphicx}
\usepackage{xcolor}

\def\ext{\textrm{e}}

\begin{document}

\title{Quantum computations with topological edge states}

\author{Igor~Timoshuk}
\affiliation{Condensed-matter physics laboratory and Physics department, HSE University, Moscow, Russia}
\author{Yuriy~Makhlin}
\affiliation{Condensed-matter physics laboratory and Physics department, HSE University, Moscow, Russia}
\affiliation{Landau Institute for theoretical physics, Chernogolovka, Russia}
\begin{abstract}
Topological quantum computations can be implemented with local Majorana zero modes. To simplify manipulations, one can use Majorana edge states in gapped two-dimensional systems. Here we demonstrate how this approach can be used for a Kitaev hexagonal model and discuss
implementation of quantum-state transfer along the edge and two-qubit gates mediated by the edge modes.
%its possible realization based on Josephson qubits.
\end{abstract}

\maketitle

%%%%%%%%%%%%%%%%%%%%%%%%%%%%%%%%%%%%%%
\section{Introduction}

Topological quantum computations~\cite{KitaevTopQC97,NayakReview,LahtinenReview} use
systems and materials, which support anyons with non-abelian statistics to perform quantum logical gates. Existence of distant non-abelian anyons ensures degeneracy of the ground state, which provides the computational quantum space. Quantum logic gates are achieved by braiding localized anyonic modes with each other to enforce a unitary operation on this degenerate ground state. Such operations are non-trivial, due to non-abelian statistics, and fault-tolerant, due to topological protection against various imperfections.
The interest to topological quantum computations, apart from their relation to fundamental properties of matter, hinges upon this tolerance to inaccurate manipulations, external noises and various kinds of disorder.

Implementation of topological quantum computations requires materials or structures, which support non-abelian anyons, as well as tools to control them. One possible candidate~\cite{WillettPRL87} is the state of the fractional Hall effect at the filling factor $\nu=5/2$. Research in this direction continues~\cite{Willett19}, however, more recently there was a surge in investigations of low-dimensional topological materials in the weak-coupling limit, including artificially constructed systems. In particular, networks of hybrid superconductor--topological insulator 1D wires, implementing 1D Kitaev model~\cite{Kitaev1Dmodel}, have been studied in detail theoretically and experimentally~\cite{Alicea10,Alicea12,LutchynNatRev18}. Further suggestions include the use of natural or artificial spin liquids, described by the Kitaev honeycomb model~\cite{KitaevHoneycomb} or its cousins. In these materials Majorana zero modes (MZMs) may be bound to defects in their structure, e.g., domain walls in 1D wires or vortices in 2D materials. In gapped materials, MZMs may be bound to the edges and exist as boundary excitations in the system.

Interest to Majorana quasiparticles in solid-state systems has grown over the recent years.
Search for Kitaev materials, which are described by the Kitaev or similar models, is underway. Recent experiments~\cite{Kasahara2018,Yokoi2021,Tanaka2022} demonstrated that $\alpha$-RuCl$_3$ may be a suitable candidate, and there is a number of further materials with similar properties, so called Kitaev materials~\cite{Trebst}. Another approach is based on design of artificial systems, mimicking Kitaev physics. For instance, one can use superconducting qubits to build a lattice with carefully designed Kitaev interactions~\cite{NoriKitaevModel,Sameti2019}. Other realizations of quantum bits can be used as well depending on the available two-qubit interactions.

Implementation of topological quantum logic operations relies on braiding of MZMs. This requires very accurate control of their position (for instance, position of vortices~\cite{IvanovMaj} or domain walls in a network of nanowires~\cite{Alicea10,Alicea12}), and in a typical realizations of such a system moving these anyons around is a challenging experimental task.
It has been suggested~\cite{Lian} that this problem may be circumvented by the use of one-dimensional edge modes, where chiral Majorana excitations move without any manipulations from outside. The use of this approach could make certain manipulations needed for the braiding operation automatic and reduce the task to the proper design of the edges~\cite{Lian,Beenakker2020}. This is especially interesting since braiding of zero-dimensional Majorana zero modes, an interesting physical phenomenon and an important step to realization of topological quantum computations, was not demonstrated so far~\cite{Beenakker2020}.

Various materials and possible realizations of Majorana edge modes were considered. Here we discuss how specifically the edge modes can be used for quantum information processing. Accordingly, we demonstrate a protocol that allows one to transfer quantum information along the edge, as a specific example we consider communication between a pair of external qubits via the edge. In related recent work transport of these modes was considered with the major goal of probing the physics of the edge MZMs. In particular, Aasen et al.~\cite{AasenPRX20} developed electrical probes of (neutral) MZMs. Klocke et al.~\cite{KlockePRL21} focused on time-domain interferometry as a means to probe the edge modes and bulk anyons and analyzed energy transport between external spins along the edge.
Feldmeier et al.~\cite{Feldmeier} propose techniques based on spin-polarized scanning tunneling microscopy to probe the charge-neutral edge states in Kitaev materials and other two-dimensional quantum magnets.
 
Motivated by these developments, we analyze possibilities to utilize the edge states in Kitaev materials for quantum-information processing (QIP) and benefit from their topological nature. More specifically, to give an example of QIP we demonstrate that a two-qubit quantum logic gate can be implemented via the edge modes between two quantum bits, coupled to the edge of a Kitaev material.

Implementation of systems with chiral edge Majorana modes may be achieved using various methods. One approach, which we focus on here, is based on using edge states of topological 2D lattice systems. These should carry edge states due to the bulk-boundary correspondence~\cite{HasanKaneReview,VolovikBook}. A well-known example is the Kitaev honeycomb model~\cite{KitaevHoneycomb}, where topologically nontrivial phase may be realized under certain conditions, and chiral Majorana edge modes emerge. In this paper, we consider such a situation and analyze, how manipulations with such edge modes may allow one to implement quantum logic gates with Majorana qubits.

While various physical systems can be considered as realizations of the Kitaev honeycomb model, and our analysis is general in this respect, in the first place here we have in mind artificial circuits, based on qubits, e.g., Josephson qubits. In these circuits, local control fields as well as spin couplings can be controlled individually, or collectively, which gives us additional control and simplifies manipulations. 
Such circuits are experimentally relevant because of the fast progress in this field. As a related development, one may mention a recent demonstration/simulation of braiding relevant for the toric-code model~\cite{GoogleToricCode} in superconducting qubit circuits.

We note that our goal is just the basic demonstration and illustration that QIP is possible with this kind of approach. Definitely, for a specific experimental implementation further questions may need to be solved, depending on the nature of qubits and further details of a particular approach. We discuss some of these questions below.

We first consider in Section~\ref{sec:edge} the edge states in the topological phase of the Kitaev model in a magnetic field. For two edge directions, the zigzag and armchair boundaries, we analyze the spectrum of the edge states and demonstrate that it contains a chiral branch in accordance with the bulk-boundary correspondence. We find the properties of the spectrum of this branch including the velocity of the edge excitations and the structure of the relevant edge states. Using these results, we discuss in Section~\ref{sec:Manip} possibilities to encode quantum information into the edge state. This information can then be transferred further along the edge, which allows one to perform quantum logic gates in this setting. In Section~\ref{sec:Disc} we summarize our findings and discuss, how various imperfections in real systems may influence properties of the edge modes and quantum-state propagation.

%%%%%%%%%%%%%%%%%%%%%%%%%%%%%%%%%%%%%%%%%%%%%%
\section{Edge modes}
\label{sec:edge}

To analyze the effects of coupling of external degrees of freedom to the edge modes, one needs to understand their properties, including the spectrum and structure of their eigenstates. Since there is little information about this in the literature (see below), we begin our analysis from the description of the edge modes for various edge directions and under various conditions.

We analyze the edge modes following the fermionization approach by Kitaev~\cite{KitaevHoneycomb}. Edge states in the Kitaev model were discussed in the original publication~\cite{KitaevHoneycomb}, including existence of the modes and qualitative features of their spectrum for the zigzag edge and in relation to to the topological properties in the 2D bulk. Analysis of the edge modes for arbitrary couplings $J_x$, $J_y$, $J_z$ in zero field (when the bulk is gapless) was performed, e.g., in Ref.~\onlinecite{Sen}. Since for the discussion of the coupling to the edge modes and their manipulation we need to know the detailed structure of the edge states and their spectrum for various edge directions, we perform this analysis below, both for the zigzag and armchair directions of the edge and under (pseudo)magnetic field, finite or vanishing at the boundary.

\begin{figure}
\includegraphics[width=0.7\columnwidth]{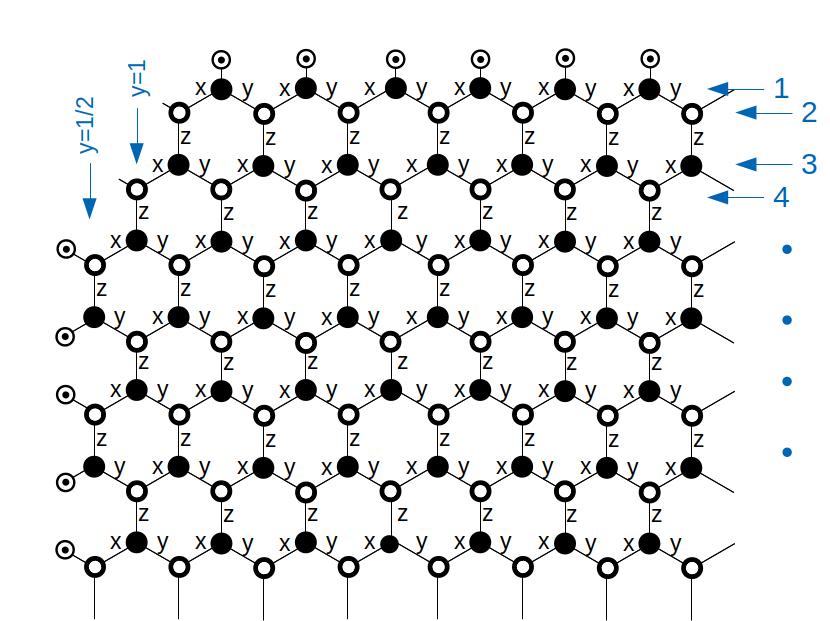}
\caption{Honeycomb lattice. $x$, $y$, $z$ mark three different directions of the links. At the zigzag edge (top) the circled dots indicate `free' $b_z$ Majorana modes. Similarly, at the armchair edge (left) the circled dots indicate `free' $b_x$ and $b_y$ Majorana modes. Arrows with numbers (right) count rows at the zigzag edge.}
\label{fig:honeyOne}
\end{figure}

We begin with a brief reminder of the Kitaev honeycomb spin model and its solution~\cite{KitaevHoneycomb}, which will be essential for our discussion below. It is defined by the Hamiltonian
\begin{eqnarray}
H &=& -J_x\sum_{x-\textrm{links}} \sigma_x^i \sigma_x^j
-J_y\sum_{y-\textrm{links}} \sigma_y^i \sigma_y^j
\nonumber\\\label{eq:fullHam}
&&-J_z\sum_{z-\textrm{links}} \sigma_z^i \sigma_z^j
-{\bf h}\sum_j \text{\boldmath$\sigma$}^j
\end{eqnarray}
with summations over links (between sites $i$, $j$) with three different directions on the honeycomb lattice, referred to as $x$-, $y$-, and $z$-links, see Fig.~\ref{fig:honeyOne}. The effect of the last Zeeman term, with summation over sites $j$, is discussed later, while first we consider the case of no magnetic field, ${\bf h}=0$.
Analysis of this model is convenient in terms of Majorana fermionic modes: on each site $i$ one defines four Majorana fermion operators, $c^i$, $b^i_{x,y,z}$, and the subspace of physical states in the whole Hilbert space is defined by additional constraints, which can be understood as fixing a Z$_2$ gauge: $D^i\equiv b^i_xb^i_yb^i_zc^i=+1$, while the spin operators are replaced by $\sigma^i_\alpha=ib_\alpha^ic^i$ with $\alpha=x,y,z$. This constraint ensures that they satisfy the standard spin algebra. Although after such fermionization the Hamiltonian appears to be of the fourth order, conservation of the relevant products $u^{jk}=i b^j_\alpha b^k_\alpha$ along all links (with $\alpha\equiv\alpha^{ij}=x,y,z$ depending on the link direction) immediately renders the Hamiltonian quadratic in each sector of fixed $u^{jk}$, with nearest-neighbor couplings $\frac{i}{2}Ju^{jk}c_jc_k$, which allows for an exact solution~\cite{KitaevHoneycomb}.

%%%%%%%%%%%%%%%%%%%%%%%%%%
%%%%%%%%%%%%%%%%%%%%%%%%%%

In the lowest-energy sector the system is translationally invariant~\cite{KitaevHoneycomb,Lieb94}, and in the momentum representation the Hamiltonian reads:
\begin{align}\label{eq:HamSqFerm}
H &= \frac{1}{2} \sum_{\bf q} A({\bf q})_{\lambda\mu}
c_{-{\bf q}\lambda} c_{{\bf q}\mu} \,,\\
A({\bf q}) &=
\begin{pmatrix}0&i f({\bf q})\\-i f(-{\bf q})&0\end{pmatrix}
\,,\label{eq:A}\\
f({\bf q}) &= 2 (J_x e^{i{\bf qn}_1} + J_y e^{i{\bf qn}_2}
+ J_z)\,,\label{eq:deff}
\end{align}
where ${\bf n}_{1,2} = (\pm1,\sqrt{3})/2$.
Here $\lambda$, $\mu$ indicate the even or odd (black or white) sublattice.
For real wave vectors $\bf q$ we have $f(-{\bf q})=f^*({\bf q})$, but the notation in Eq.~(\ref{eq:deff}) allows one to consider also complex momenta, which will be relevant near the edge.
The resulting excitation spectrum is
\begin{equation}
\varepsilon({\bf q}) = |f({\bf q})| \,.
\end{equation}

Depending on the values of the coupling constants $J_{x,y,z}$ various phases can be realized~\cite{KitaevHoneycomb}. If they satisfy the triangle inequality ($|J_x|<|J_y+J_z|$, $|J_y|<|J_x+J_z|$, $|J_z|<|J_x+J_y|$), the system is in a gapless $B$-phase, which will be of interest to us below. In this case, the gap in the spectrum closes at two opposite values of momentum, $\pm {\bf q}^*$, in the Brillouin zone.
The existence of these nodes is topologically protected by time-reversal symmetry (since under time reversal the structure of (\ref{eq:A}) persists). Below we assume that $J_{x,y,z}$ are in this range, and for most quantitative estimates that they are equal, where this does not change the situation qualitatively.

We are interested in a situation with a gapful 2D bulk. The gap can be opened by breaking the time-reversal symmetry with a (pseudo-)magnetic field, the last term in Eq.~(\ref{eq:fullHam}) (its physical nature depends on a specific realization of Kitaev model).
In a weak field, $h\ll J$ the effect of the field is described, perturbatively, by the third-order contribution:
\begin{equation}\label{eq:kappa}
V^{(3)} = -\kappa \sum_{jkl} \sigma_x^j\sigma_y^k\sigma_z^l \,,
\end{equation}
where summation is performed over triples $jkl$, in which one site is connected with the other two~\cite{KitaevHoneycomb}.
In this case, $\kappa\propto h^3$; in general, if other, weak symmetry-conserving interactions are taken into account~\cite{AasenPRX20,Balents2016}, 
$\kappa$ is linear in $h$ with a small prefactor and anisotropic in general, cf.~Section~\ref{sec:Disc} for more details. Thus, we obtain Majorana fermions on a honeycomb lattice with nearest- and next-nearest-neighbor couplings ($J$- and $\kappa$-terms), cf. Eq.~(48) in Ref.~\onlinecite{KitaevHoneycomb}.

This term~(\ref{eq:kappa}) also reduces to a term, quadratic in fermions, which couples next-nearest neighbors, $\frac{i}{2}\kappa c_jc_l$, and the updated Hamiltonian (\ref{eq:HamSqFerm}) involves the matrix
\begin{eqnarray}\label{eq:Deltaf}
A({\bf q}) &=
\begin{pmatrix}\Delta({\bf q})&i f({\bf q})\\
-i f(-{\bf q})&-\Delta({\bf q})\end{pmatrix}
\,.
\end{eqnarray}
Here $\Delta({\bf q}) = 4\kappa [\sin({\bf qn}_1)
+ \sin(-{\bf qn}_2) + \sin({\bf q}({\bf n}_2-{\bf n}_1))]$.
Near the nodes $\pm{\bf q}^*$ of the spectrum it reduces to
\begin{equation}\label{eq:nodespec}
\varepsilon({\bf q}) \approx
\pm\sqrt{3J^2\delta{\bf q}^2 + \Delta^2} \,,
\ \delta{\bf q}={\bf q} \mp {\bf q}^* \,,
\Delta=6\sqrt{3}\kappa \,.
\end{equation}

Thus the spectrum in the 2D bulk is gapped, without excitations at sufficiently low temperatures. However, topological considerations~\cite{HasanKaneReview,VolovikBook} guarantee the bulk-boundary correspondence, which implies a fermionic zero mode at the boundary. To verify this and find the spectrum of this mode, we consider the Hamiltonian near the edge and find properties of the edge mode needed for quantitative analysis of manipulations of its quantum state.

To compare various options, we consider several configurations at the edge and generalize the results of Ref.~\onlinecite{KitaevHoneycomb}. These include the zigzag and the armchair edges. Furthermore, for each edge direction we analyze the sub-cases of a uniform $h$-field applied either to all sites or to all sites except at the boundary, since this case is of interest to us, cf.~next section; for implementation in qubit networks, realization of this scenario is similar in complexity to the case of a uniform field, and thus the field at the boundary sites can be considered as an independent parameter $h_b$.
We describe these cases in detail below.

%%%%%%%%%%%%%%%%%%%%%%%%%%
\begin{figure}	\includegraphics[width=\columnwidth]{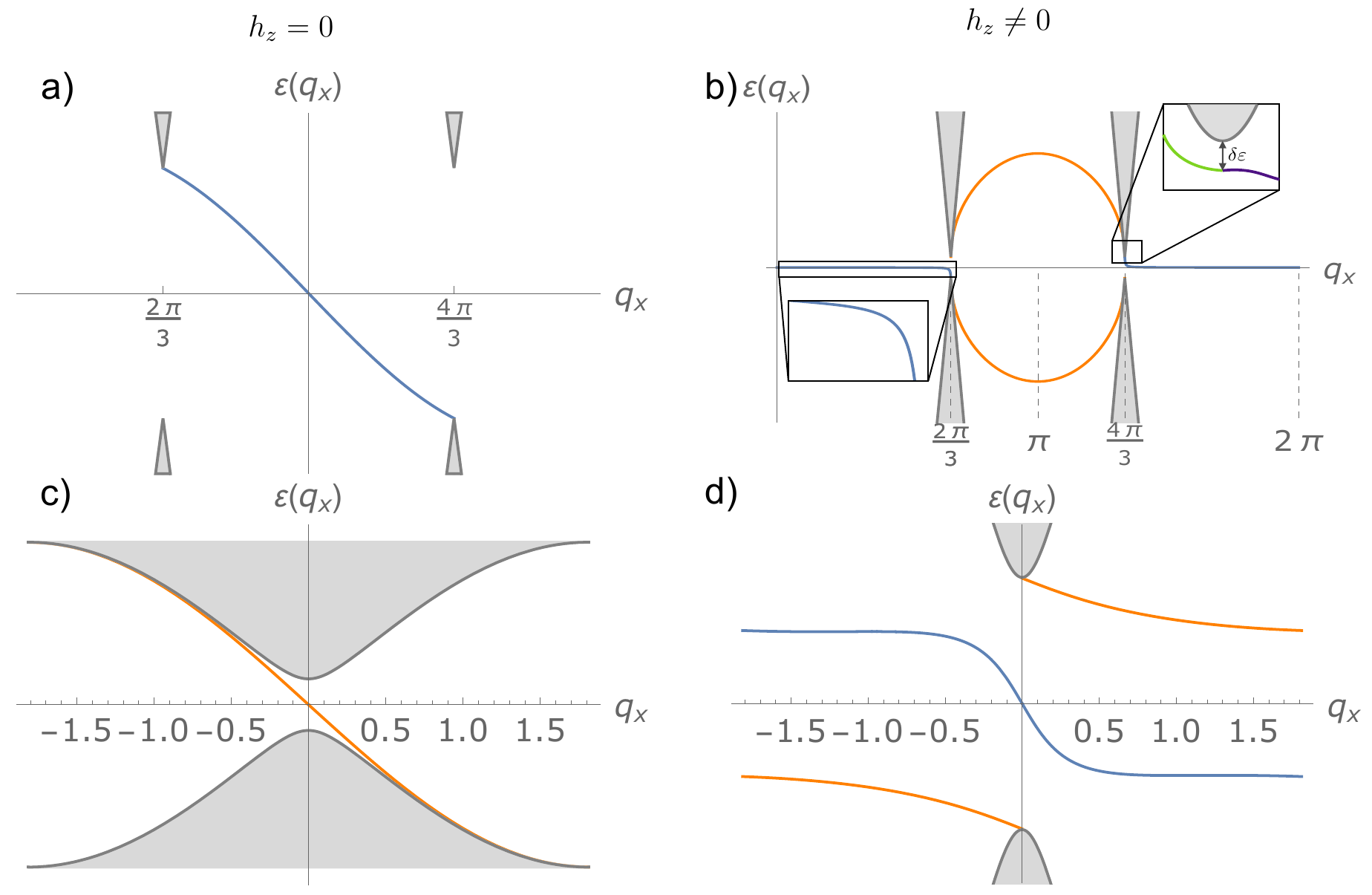}
\caption{Summary of our results for the edge spectrum for two edge directions and various patterns of pseudo-magnetic field.
The zigzag edge with (a) no field at the boundary and (b) a uniform field, incl. at the boundary.
Two regions of the continuous spectrum (grey area) nearly touch the zero energy near $q_x^*=\pm2\pi/3$. The edge-mode spectrum is shown in orange and blue in two ranges of momentum $q_x$  along the edge. It crosses zero at $q_x=0$ at a finite slope $v_\textrm{gr}$, see text.
Near the nodes at $\pm q_x^*$ the edge-mode spectrum is close to the continuous spectrum.
The spectrum for the armchair edge is shown in (c) for zero field at the boundary and (d) for a uniform field.
The continuous spectrum nearly touches zero energy close to $q_x^*=0$, the projection of both spectral nodes in the bulk onto the edge direction. The width of the Brillouin zone is $2\pi/\sqrt{3}$.
}
\label{fig:Sp4}
\end{figure}
%%%%%%%%%%%%%%%%%%%%%%%%%%
%%%%%%%%%%%%%%%%%%%%%%%%%%
\begin{figure}	\includegraphics[width=0.75\columnwidth]{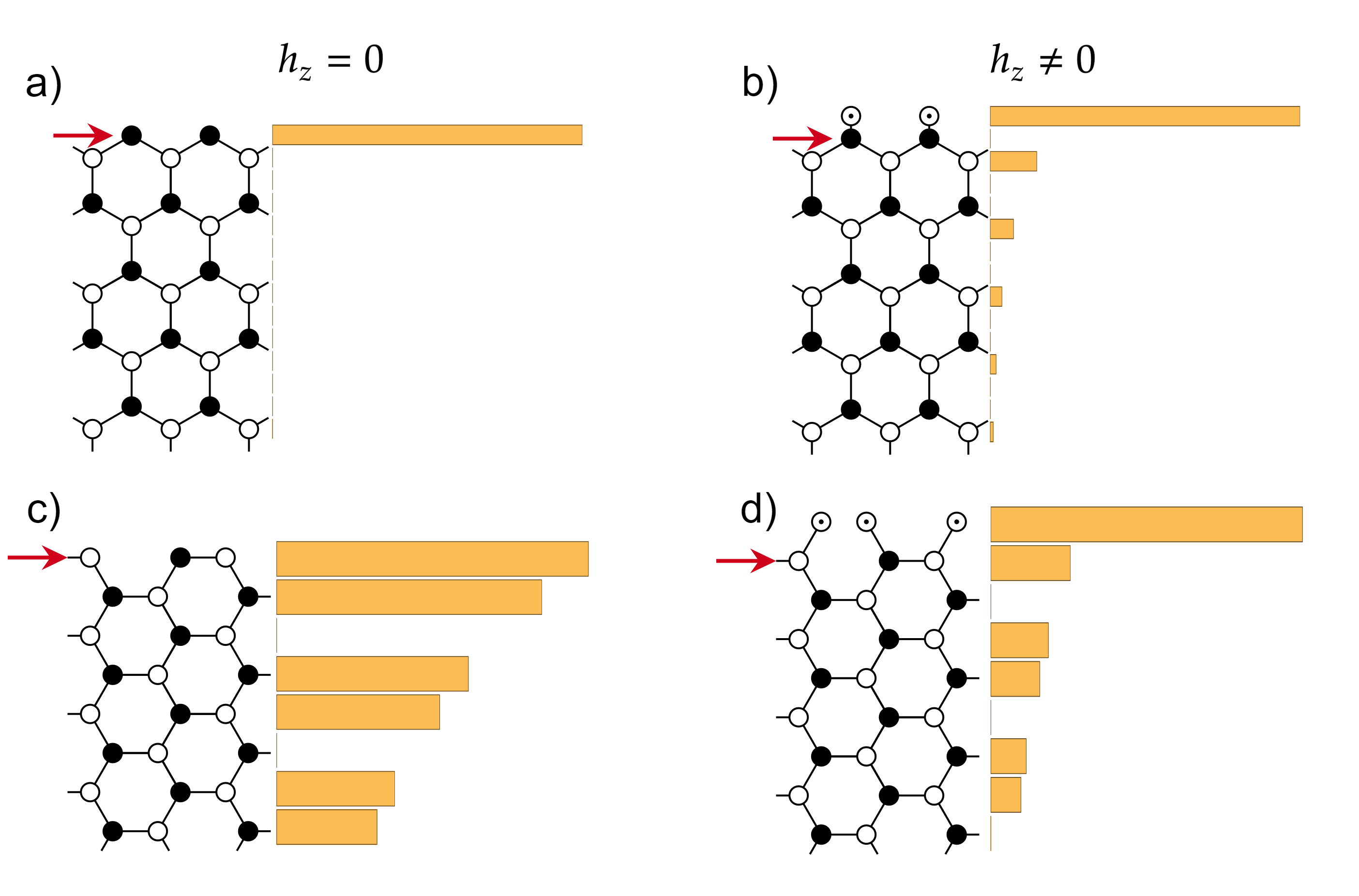}	\caption{Qualitative behaviour of the eigenmodes of the edge spectrum away from the edge (top down) for the zigzag (top row) and armchair (bottom row) edges.
We consider the cases of a uniform pseudomagnetic field right column) and the field, vanishing at the boundary (left column).}
\label{fig:wf4}
\end{figure}
%%%%%%%%%%%%%%%%%%%%%%%%%%

Near the edge, the system is translationally invariant only along the edge.
Furthermore, under a magnetic field, in the fermionic language in addition to the next-nearest $\kappa$-couplings from (\ref{eq:kappa}), one should keep the local $b$-$c$ coupling at the edge sites with a strength given by the field $h_b$ at the boundary.
We take this term into account first, and then include the weaker $\kappa$-term. In our analysis we focus on the situation, where the three components of the field $h_x$, $h_y$, $h_z$ are generally of the same order. Then for the pure Kitaev model, $\kappa\propto h^3$ (see above and Ref.~\onlinecite{KitaevHoneycomb}).
In general, especially in Kitaev materials~\cite{Trebst}, other, weak symmetry-conserving interactions may be relevant. Taking them into account~\cite{AasenPRX20,Balents2016}, makes 
$\kappa$ linear in $h$ with a small prefactor and anisotropic, cf.~Section~\ref{sec:Disc} for more details.
In our analysis we consider these two terms in the Hamiltonian as independent and refer to them as $h$- and $\kappa$-terms (or $\propto h$, $\propto\kappa$ terms, respectively).

We begin with a summary for the zigzag edge direction (Fig.~\ref{fig:honeyOne}, top edge), analyzed in Ref.~\onlinecite{KitaevHoneycomb}. For a given momentum $q_x$ along the edge, using the transverse row numbers, see Fig.~\ref{fig:honeyOne}, we perform perturbative analysis in the weak field.
The resulting spectrum is shown in Fig.~{\ref{fig:Sp4}b.
The continuous gray areas near $\pm q_x^*=\pm 2\pi/3$ are projections of the bulk spectrum near $\pm{\bf q}^*$ onto the edge. For $h=0$ one finds a flat zero-energy mode in the range $-2\pi/3 < q_x <2\pi/3$ and two flat zero-energy modes in the remaining range of $2\pi/3 <q_x< 4\pi/3$. In the two-mode region, the $h$-term splits the  spectrum, $\varepsilon=\pm 2h_z\sqrt{1-4\cos^2(q_x/2)}$, and the $\kappa$-term has little further effect.
In contrast, in the single-mode range, the $h$-term does not change the energy but modifies the wave function of the edge mode, giving it an order-$h$ admixture of the sites away from the edge (this admixture decays on the scale of a few sites, depending on $q_x$). The $\kappa$-term, after averaging over this eigenmode, leads to the spectrum
\begin{equation}\label{eq:SpSingleMode}
\varepsilon_0(q_x) = -\frac{h_z^2\kappa}{J^2}
\frac{\sin q_x + \tan\frac{q_x}{2}}%
{\cos^2\frac{q_x}{2} -\frac{1}{4} +\frac{h_z^2}{4J^2}}
\,.
\end{equation}
This expression applies everywhere except in the $\sim\Delta/J$ vicinity of $\pm2\pi/3$, where it gives energies comparable to the bulk gap. Thus the spectrum crosses zero at $q_x=0$ with a weak slope, which indicates the group velocity of
\begin{equation}\label{eq:vgrzigzag}
v_\textrm{gr} = - \frac{h_z^2\kappa}{2J^2} \,,
\end{equation}
which is of high order in the field.
Near $q_x=\pm 2\pi/3$ this spectrum nearly touches the continuum, but in a narrow range $\delta q_x\lesssim \Delta/J$ around this point higher-order corrections due to close proximity to the continuous spectrum become relevant.

If the magnetic field is applied everywhere except for the edge row (row 1 in Fig.~\ref{fig:honeyOne}), $h_b=0$, the spectrum is modified considerably. The outermost $b_z$-mode decouples from the rest of the sample, and comprises a flat zero-energy mode. At the same time, another edge mode appears in the range $2\pi/3<q_x<4\pi/3$ and crosses zero at $q_x=\pi$.
To be specific, we assume that only the $h_z$ component vanishes at the edge so that the $\kappa$-couplings near the edge are modified as little as possible.
Then essentially, the spectrum of this mode can be found in Ref.~\onlinecite{KitaevHoneycomb}, where it appeared at an intermediate step of the discussion of the edge spectrum:
\begin{equation}\label{eq:epszigzagh=0}
\varepsilon = -12\kappa\sin q_x \,,
\end{equation}
with the velocity $v_\textrm{gr}=-12\kappa$, see Fig.~\ref{fig:Sp4}a. One should remark, however, that strictly speaking in contrast to the calculation in Ref.~\onlinecite{KitaevHoneycomb}, vanishing $h_z$ at the zigzag edge also suppresses the $\kappa$-terms within row 2, see Fig.~\ref{fig:honeyOne}, but this affects the edge spectrum only weakly, in higher orders in $\kappa$.
Analysis of the spatial structure of the corresponding mode shows that it is localized on a few sites near the edge and only in the odd rows, in contrast to the case of a finite field at the boundary as considered above. Hence coupling of an external spin in this configuration may be considerably stronger.

At the armchair boundary (Fig.~\ref{fig:honeyOne}, left edge) the spectrum should have the same topological properties but its behavior differs from the above, cf.~Fig.~\ref{fig:Sp4}c,d.
For this direction of the boundary the projections of two bulk spectral nodes with conical spectrum coincide, and there is only one point, $q_x^*=0$, where the continuous spectrum touches zero energy at zero field ($x$ is chosen along the edge).
We first consider the case of a weak magnetic field in the bulk with no field, $h_b=0$, applied at the very edge (at the sites with `free' $b_x$, $b_y$ in Fig.~\ref{fig:honeyOne}). In this case one can construct the edge mode from the plane-wave states in the bulk, so that the combination satisfies the boundary conditions.
We show below that the eigenmode at $q_x=0$ behaves approximately as
\begin{equation}\label{eq:armchairmode}
\begin{pmatrix}1\\-1\end{pmatrix}
\sin\frac{4\pi y}{3} e^{-|\Delta| y/\sqrt{3}J} \,,
\end{equation}
where the positive direction of $y$ is chosen from the edge towards the bulk, and the spinor notation refers to the black/white sublattices, see Fig.~\ref{fig:Scheme}. The (cartesian) coordinate $y$ is measured from the ``first non-existent'' row (just outside the edge), that is the eigenmode vanishes at that row, while the outermost row has $y=1/2$, see Fig.~\ref{fig:honeyOne}.
This zero-energy eigenmode is constructed as a combination $C_+\psi_++C_-\psi_-$ of two plane-wave bulk solutions, $\psi_\pm= (1,-1)^T
e^{i{\bf q}_\pm{\bf r}}$, with complex momenta ${\bf q}_\pm = \pm{\bf q}^*+i\delta q\hat y$ near the nodes $\pm{\bf q}^*$ of the bulk spectrum. At $\delta q=\Delta/\sqrt{3}J$ it is a solution with zero energy (\ref{eq:nodespec}), that is $[H,\psi]=0$. It cannot be normalized in the bulk sample, but when considered in a half-plane $y>0$, it is normalizable and, importantly, the zero-energy condition $[H,\psi]=0$ is satisfied at all sites, including those at the edge: indeed, the cross-edge $J$-coupling to the row $y=0$ with vanishing $\psi$ in a bulk sample is replaced by the equivalent absence of coupling to this `outside' row beyond the edge.
This also applies to the $\kappa$-couplings, except the cross-border terms that couple the rows $y=\pm1/2$.
%The imaginary momentum is allowed near the edge and is
%responsible for the exponential decay to the bulk.
%One can verify that this is a relevant eigenmode by comparing %the Hamiltonian of the half-plane with the full plane (e.g., 
%with antisymmetric conditions beyond the edge $y=0$).
Notably, these latter terms can be treated as a perturbation, they are proportional to the $\sigma_z$ Pauli matrix in the sublattice space. This perturbation (i) does not change the energy at $q_x=0$ and (ii) at other $q_x$ modifies it only weakly due to the weak $\kappa$-coupling and a weak weight at the edge due to large penetration depth $\sqrt{3}J/\kappa$ in Eq.~(\ref{eq:armchairmode}). This gives only a minor correction to the group velocity of the edge mode. To find the group velocity, one considers the perturbation of the Hamiltonian due to finite $q_x$, which modifies the Hamiltonian (\ref{eq:Deltaf}) by $\delta f \approx i\sqrt{3}Jq_x$. Thus we find $v_\textrm{gr}=-\sqrt{3}J$, corresponding to the velocity of the zero mode near the nodes in the bulk. Numerical results for the spectrum in Fig.~\ref{fig:Sp4}c illustrate this behaviour.

Finally, we consider applying a finite field $h_x=h_y=h_b$ also to the edge sites at the armchair edge (we chose $h_x=h_y$ just for illustration). To analyze the modification of the spectrum, we recall that at $h_b=0$ on top of the aforementioned zero mode, two extra modes should be considered: these correspond to the `free' sites of $b_x$ and $b_y$ (pertaining to edge spins in the spin model) and have flat-band zero-energy spectra at all $q_x$ since they are completely decoupled from the rest of the system at $h_b=0$. A finite value of the edge field $h_b$ lifts the degeneracy of these three zero modes by coupling the $b$-modes to the $c-$mode in Eq.~(\ref{eq:armchairmode}). In addition, the second-order terms due to coupling to the continuous spectrum produce a finite slope $\sim h_b^2$ of the remaining zero mode (however, one can also manipulate the other low-energy modes at low $h_b$). To find the velocity (slope) of the single remaining zero mode, we first note that the wave function of this mode, which at zero edge field $h_b=0$ is localized at the `free' $b$-sites at the boundary is modified by $h_b$. Indeed, as above in the derivation of (\ref{eq:armchairmode}) one can construct the $h_b$-perturbed wave function from the solutions near the nodes $\pm{\bf q}^*$ with an imaginary momentum. However, a finite $h_b$ modifies the boundary conditions at the edge. Indeed,
now we look for a zero mode $C_xb_x+C_yb_y+C_+\psi_+ + C_-\psi_-$.
From the condition that its commutator with $H$ vanishes at the $b_{x,y}$ sites, one finds that $C_+\psi_+ +C_-\psi_-$ vanishes at the edge row, $y=1/2$. Further, vanishing of $[H,\psi]$ at the edge row $y=1/2$ implies $h_bC_x=h_bC_y=-J\psi_\uparrow(y=1)$. Solving these relations, we find the zero mode
\begin{equation}\label{eq:armhmode}
\begin{pmatrix}1\\1\end{pmatrix}_b
-
\frac{2h_b}{\sqrt{3}J}
\begin{pmatrix}1\\-1\end{pmatrix}
\sin \frac{4\pi y-2\pi}{3}
e^{-|\Delta| y/\sqrt{3}J}
\,,
\end{equation}
where the subscript $b$ indicates the value of the eigenmode at the `free' $b$-sites $C_x=C_y=1$ (the upper and lower values correspond to $b_x$ and $b_y$) and the oscillating part is shifted compared to (\ref{eq:armchairmode}) and vanishes at the outermost edge sites, where $y=1/2$.

We find the velocity of this mode by again considering the energy at finite $q_x$ due to the perturbation $\delta f \approx i\sqrt{3}Jq_x$ for the bulk sites. After averaging over the eigenmode (\ref{eq:armhmode}), this produces a weighted average of the group velocity in the bulk, $-\sqrt{3}J$, and the vanishing velocity at the edge. We find
\begin{equation}
v_\textrm{gr} = -\sqrt{3}{J} \frac{2h_b^2}{2h_b^2+\sqrt{3}|\Delta|J}
\,,
\end{equation}
in agreement with direct simulations, cf.~Fig.\ref{fig:Sp4}d.
As a result, the velocity of the zero edge mode scales at weak edge fields as $v_\textrm{gr}\sim - h_b^2/|Delta|$, grows with $h_b$ and saturates in stronger fields of order $\sqrt{|\Delta|J}$ at the bulk value $-\sqrt{3}J$.
The spectrum at an intermediate $h_b$ is illustrated in Fig.~\ref{fig:Sp4}d. The energy of the edge modes away from $q_x=0$ appears in the second order in $h_b$ due to the coupling to the continuous spectrum and is of order $\pm h_b^2/J$.

Thus, we find that the spectrum of an armchair-edge mode crosses zero energy exactly at the point $q^*_x=0$ near the minimum of the continuous spectrum. This has important consequences, relevant for manipulations of the edge states discussed in Section~\ref{sec:Manip}. 
Indeed, the edge state penetrates into the bulk, in the direction transverse to the edge, much deeper, with penetration depth $\sim J/\Delta$. Hence its weight at the edge is suppressed.

Fig.~\ref{fig:wf4} illustrates qualitative behavior of the eigenmodes in the four cases considered.

%%%%%%%%%%%%%%%%%%%%%%%%%%%%%%%%%%%%%%%%%%%%%
\section{Operations with edge modes}
\label{sec:Manip}

In this section, we analyze influence of external perturbations on the state and evolution of the edge modes. This also allows us to propose and discuss possibilities to implement quantum logic gates, based on qubits formed by such modes.
We focus on the zigzag edge, since it appears most suitable for such manipulations. However, we also discuss extensions of these procedures to the armchair, or any other edge with the goal to compare their suitability for the needed manipulations.

We discuss manipulations with the quantum state of the edge modes using two approaches. First, one can apply a local (pseudo-)magnetic field $\delta h$ to a single spin at the edge, or to a number of spins, in order to influence the quantum state. This adds an extra term $-\sum_i \delta h^i_z(t) \sigma^i_z$ to the Hamiltonian. Here summation is over the edge sites, the coefficients $\delta h^i_z(t)$ indicate the spatial and temporal profile of the additional applied field, and we assume that the field is in the $z$-direction (although other directions can also be considered, the description is more straightforward in this case). It is convenient to describe the effect of the edge field in the Majorana language, in which the extra term is of the type $i \sum_i \delta h_z^i c^i b^i_z$.
We can find its effect from perturbative analysis, but it can also be deduced from the discussion in Section~\ref{sec:edge}, where $h_z$ played the part of the field at the edge. Thus, we arrive at the conclusion that in accordance with Eqs.~(\ref{eq:vgrzigzag}), (\ref{eq:epszigzagh=0}) a field at the boundary with a sufficiently smooth profile modifies the local velocity of the edge mode, without otherwise affecting its quantum state. Moreover, even a sharper profile could induce only forward scattering, which also can be looked at as a modification of the velocity of the edge mode (or its travel time).
This gives one a useful tool of manipulation of the traveling wave packets at the edge, which can be used in all kinds of interferometric experiments and for controlled quantum-information transfer along the edge.
Note that both for the zigzag and armchair edge, the value of the field at the boundary controls the velocity (and also the structure) of the edge mode. This can be used to choose a convenient range of velocities for the experiment as well as to control the velocity {\em in situ}.

Alternatively, one can couple, in a controlled manner, the edge to an external spin/qubit (which can be viewed as a `quantum external field') in order to perform a joint quantum gate on the edge and the external qubit. In particular, this may allow for a SWAP operation, which exchanges the states of the external qubit and the edge, see below for a detailed discussion.

Consider coupling an external spin $\sigma^\ext$ to a site $i$ at the edge of the system in a similar fashion, $\lambda\sigma_z^\ext \sigma^i_z$. To describe the effect of the field, it is convenient to use the Majorana representation. One may consider the external spin as a part of the system with the Kitaev Hamiltonian, since the $zz$-coupling to the edge spins has not been used. Thus, introducing four extra Majorana operators $c^\ext$, $b_{x,y,z}^\ext$ pertaining to the external spin, one may represent $\sigma^\ext_z$ by $ib^\ext_z c^\ext$. As above, this reduces the Hamiltonian to a quadratic combination of $c$-operators. The link operator $u^\ext=ib^\ext_z b^i_z$ commutes with the Hamiltonian (it can be considered as one of the link operators in the Kitaev model on the sample augmented by the external site). This reduces the coupling term to $iu^\ext\lambda c^\ext\psi^i$. This coupling, if switched on for a properly chosen finite period and neglecting the velocity of the edge mode, could produce the exchange/swap of the two Majorana modes, which is equivalent to the braiding operation: for instance, the evolution operator $\exp(-i\cdot(i\frac{\pi}{4} c^\ext \psi^i))$ maps $c^\ext\to -\psi^i$ and $\psi^i\to c^\ext$ (here we replaced the link operator $u^\ext$ by 1, see, however, discussion at the end of this section and in Appendix~\ref{app:SWAP}). This effectively writes the $c^\ext$-mode onto the edge~\cite{Vishveshwara}.
This operation, with duration $\sim 1/\lambda$, involves only a single edge site if the operation is fast enough, $\lambda\gg v_{\textrm{gr}}$. On the other hand, if we keep $\lambda$ below the gap value to the continuous spectrum in the relevant range of momenta, the operation is adiabatic with respect to these higher states, and they are not excited. These conditions are consistent for the zigzag edge mode. They can also be consistent at the armchair edge at weak edge fields. If, for some edge structure, they are inconsistent, the $c^\ext$-mode is swapped onto a wider wave packet at the edge, and further analysis is required to optimize the applied pulse $\lambda(t)$ and evaluate the quantum operation performed, see also Section~\ref{sec:Disc}.

Now we apply these general considerations to the specific spectrum and structure of the considered edge modes. This includes the edge mode at the zigzag or armchair edge near $q_x=0$. We further consider the case of a system subject to magnetic field in the bulk, but not at the edge sites, since in this case a stronger coupling to the edge is possible as we find below.

For the armchair edge, due to the proximity of the continuous spectrum, the zero-mode wave function has a large penetration depth, $\sim J/\Delta$, and its weight at the edge is relatively weak $\sim \Delta/J$, which strongly suppresses the coupling to the external spin/mode. Thus, the use of the armchair edge is possible but inconvenient.

For the zigzag edge the wave function of the edge mode near $q_x=0$ is mostly localized at the edge $b_z$-mode, its amplitude in the bulk is proportional to $h_z/J$ and decays into the bulk much faster, than for the armchair edge above, with the decay factor $(2\cos(q_x/2))^{-1}$. However, its weight on the $c^i$-modes is non-zero only on the even rows from the edge (empty circles in Fig.~\ref{fig:honeyOne}), and vanishes in the outermost row, see~Fig.~\ref{fig:wf4}. Hence, coupling of the external site to the edge would be vanishingly weak.

Thus, on one hand, the zigzag edge appears more promising since in this case the edge state is better localized near the boundary and has a higher weight there, but on the other hand, in the considered configuration this weight vanishes at the outermost coupling site, see~Fig.~\ref{fig:wf4}c. One possibility to increase the coupling would be to couple the external qubit to a site not in the outermost row but in the next (even) row with an appreciable weight of the edge state. However, in this case the extra coupling term does not immediately reduce to a quadratic form since the second-row sites are already coupled via all kinds of links, $xx$, $yy$, and $zz$,
and analytical description and effects of this coupling are more complicated. Here we suggest the following alternative: instead of applying the magnetic field uniformly in the bulk, one may apply it everywhere except for the edge row (row 1 in Fig.~\ref{fig:honeyOne}). This would allow for a stronger coupling and faster operation.

The described operation between the external spin and the edge can be extended to operations between two external spins. We describe in Appendix~\ref{app:SWAP} the SWAP operation on these spins. In particular, we also discuss there dependence of the operations above on the link operator $u^\ext$, which appears to indicate undesired entanglement between the edge and external spin. We show that the extended version allows for two-qubit gates, without entanglement with the edge.

We further note that modification of this procedure allows one to perform arbitrary two-qubit gates between the external spins via the edge modes; this can be seen from the fact that spin-edge two-qubit gates together with local single-qubit gates (performed via local fields) form a universal gate set~\cite{DVDUniv95,DeutschUniv95,9authorsUniv95} for quantum computations (this is illustrated in Appendix~\ref{app:SWAP} for the SWAP gate in the fermionic language). This observation is of interest since the SWAP operation itself is not a universal two-qubit gate. It implies that the edge mode can be used as a mediator in universal quantum computations.

%%%%%%%%%%%%%%%%%%%%%%%%%%%%%%%%%%%%%%%%%%%%
\section{Discussion}
\label{sec:Disc}

Our analysis indicates that strong coupling to the edge is most conveniently achieved at the zigzag edge. Using the methods of the coupling described above and the possibilities to write states of external modes onto the edge as well as to read it out by a similar method, one can consider various extensions. For instance, if two qubits are coupled to the edge at different points, Fig.~\ref{fig:Schemes}a, one can entangle them via the edge by first writing their states onto the edge, using the edge to deliver each state to the other qubit and then reading these out, as described in Appendix~\ref{app:SWAP}.

Another possibility is to couple edges of different samples with each other. For instance, Fig.~\ref{fig:Schemes}b shows coupled samples of the Kitaev model with different Chern numbers (+1 and -1). In this case one can effect braiding of the Majorana edge modes, as the figure indicates, which may be considered as a quantum logic gate~\cite{Lian}. Fig.~\ref{fig:Schemes}c shows an extension of this approach towards a larger-scale quantum-coherent device. Each hexagonal cell here is a piece of Kitaev material and can be considered as a qubit, which is coupled to its neighbors at edges via controlled couplings (indicated by red dots). Proper operation of the device should allow for quantum-computing operations.

Let us also comment on further aspects of the operation. First, the packet created by coupling the edge to an external spin/qubit may be distorted during propagation along the edge because of the nonlinearity in the spectrum. To suppress this distortion one may create a wider packet of the lowest-energy edge excitations, see below.
On the other hand, one can show that nonlinearity in the edge-mode spectrum may be suppressed by a carefully tuned profile of the $h$-field in the direction transverse to the edge; this, however, may be challenging experimentally.

A related aspect concerns the procedure for writing an external Majorana to the edge (and the inverse readout procedure). In Section~\ref{sec:Manip} we discussed a point-like coupling of intermediate strength, which allows for fast and local Majorana swap onto or from the edge. Alternatives to this procedure may be considered. One can couple strongly to the edge, with $\lambda>J$, to enable even faster Majorana swap. However,
this would couple to the bulk spectrum, degrading fidelity of the local swap, and a point-like coupling would create a narrow wavepacket subject to strong distortion during propagation since the edge spectrum is not perfectly linear. To enable creation of wider wave packets, one may employ coupling not to a single, but to a range of edge spins. An algorithm was proposed~\cite{TTM}, which enables creation of a wide wave packet with full control of its shape. In particular, one can create a shape with no overlap with the bulk states.
Nevertheless, numerical simulations of the evolution show that operation with reasonably high fidelity of the quantum gate can be achieved even without these steps. With a point-like weak coupling, $\lambda\ll v_\textrm{gr}$, of an external qubit to the edge in numerical simulations without further optimization we obtained fidelity higher than 96\% of the state transfer between external qubits. Propagation of the created wave packet is shown in Fig.~\ref{fig:simul}. Thus, one can choose and optimize a suitable read/write method depending on particular realization of the Kitaev-honeycomb system and further limitations.

%%%%%%%%%%%%%%%%%%%%%%%%%%
\begin{figure}
\begin{tabular}{ccc}
\includegraphics[width=0.3\columnwidth]{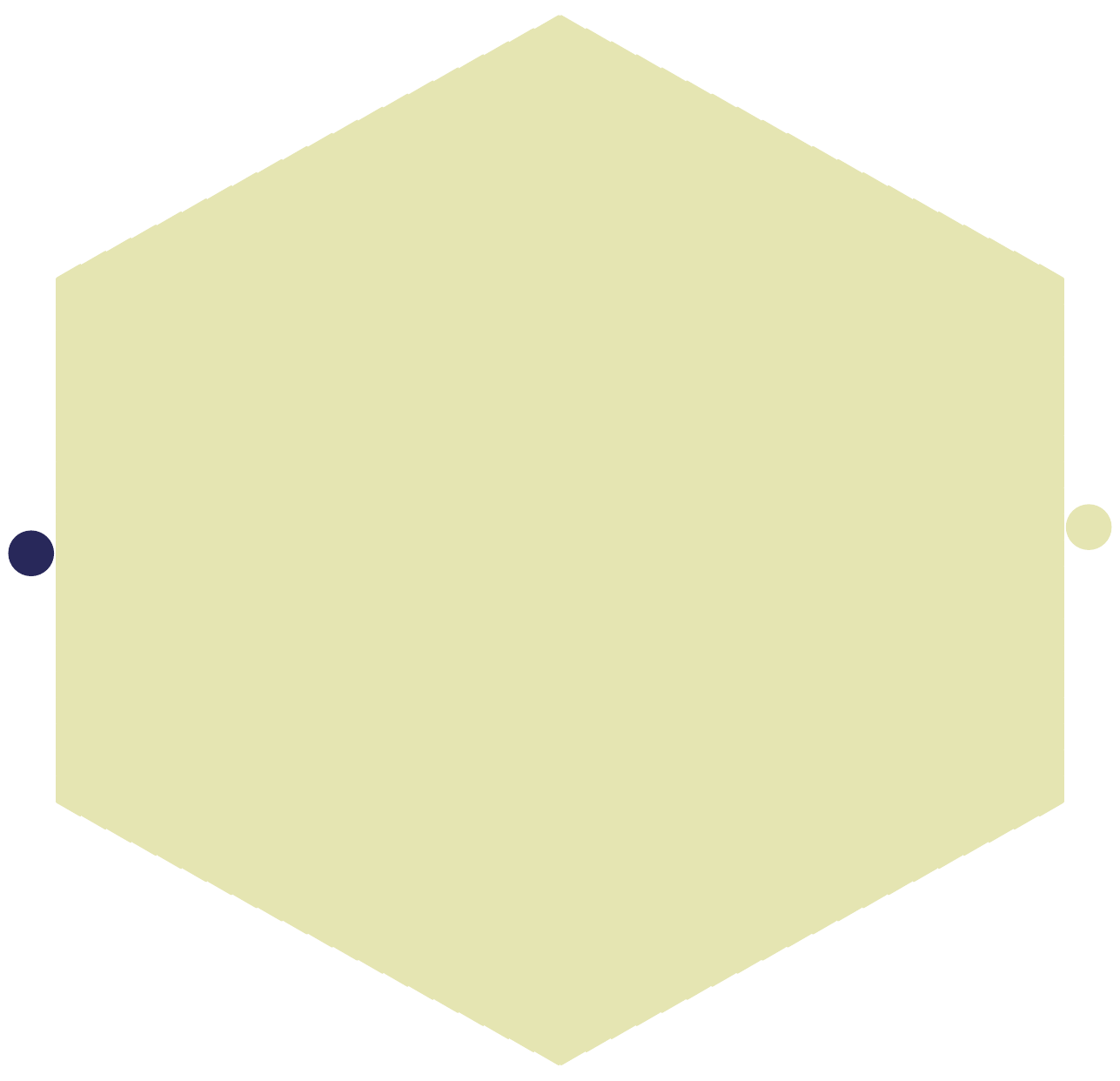}
&\includegraphics[width=0.3\columnwidth]{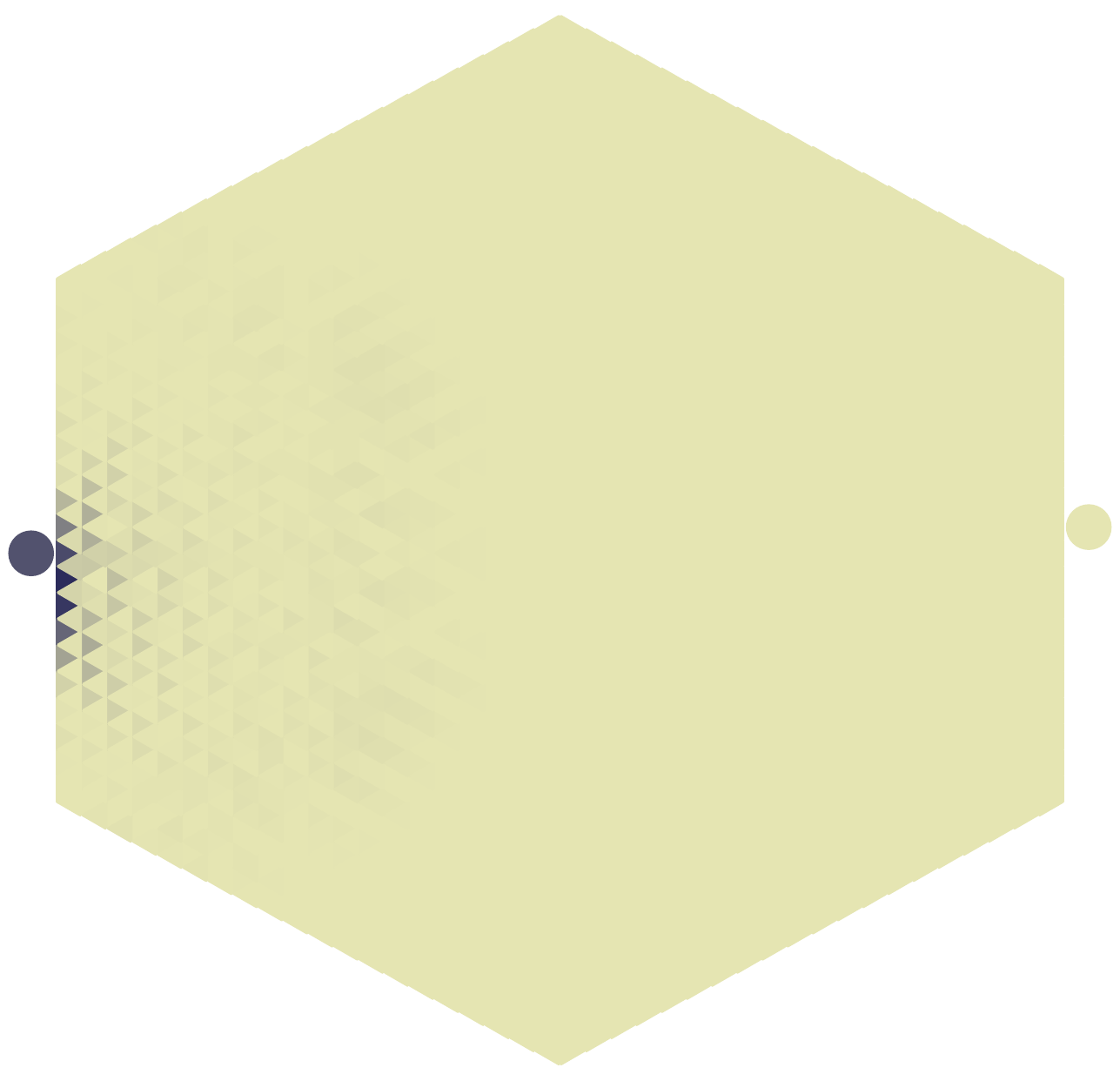}
&\includegraphics[width=0.3\columnwidth]{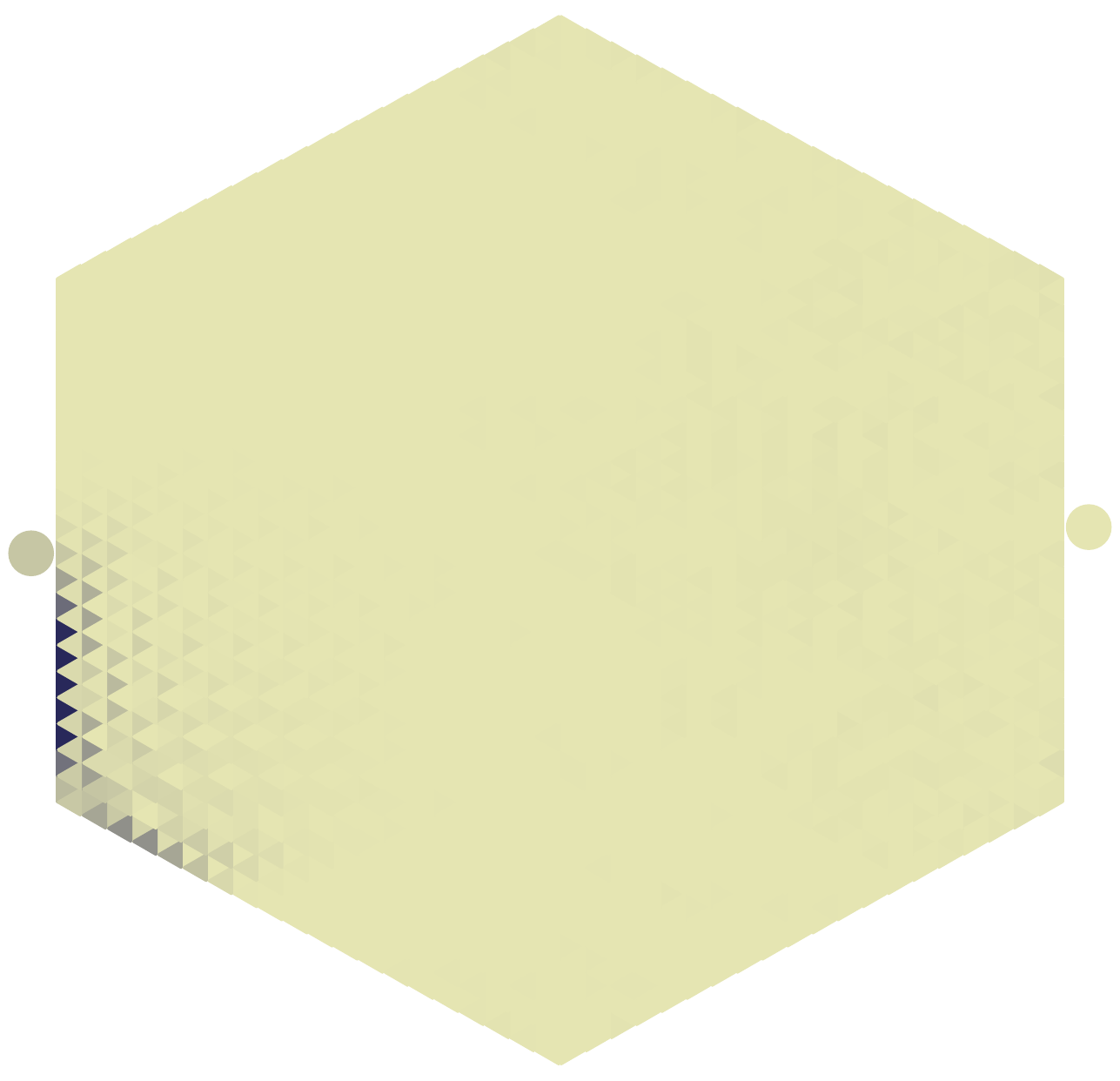}
\\
\includegraphics[width=0.3\columnwidth]{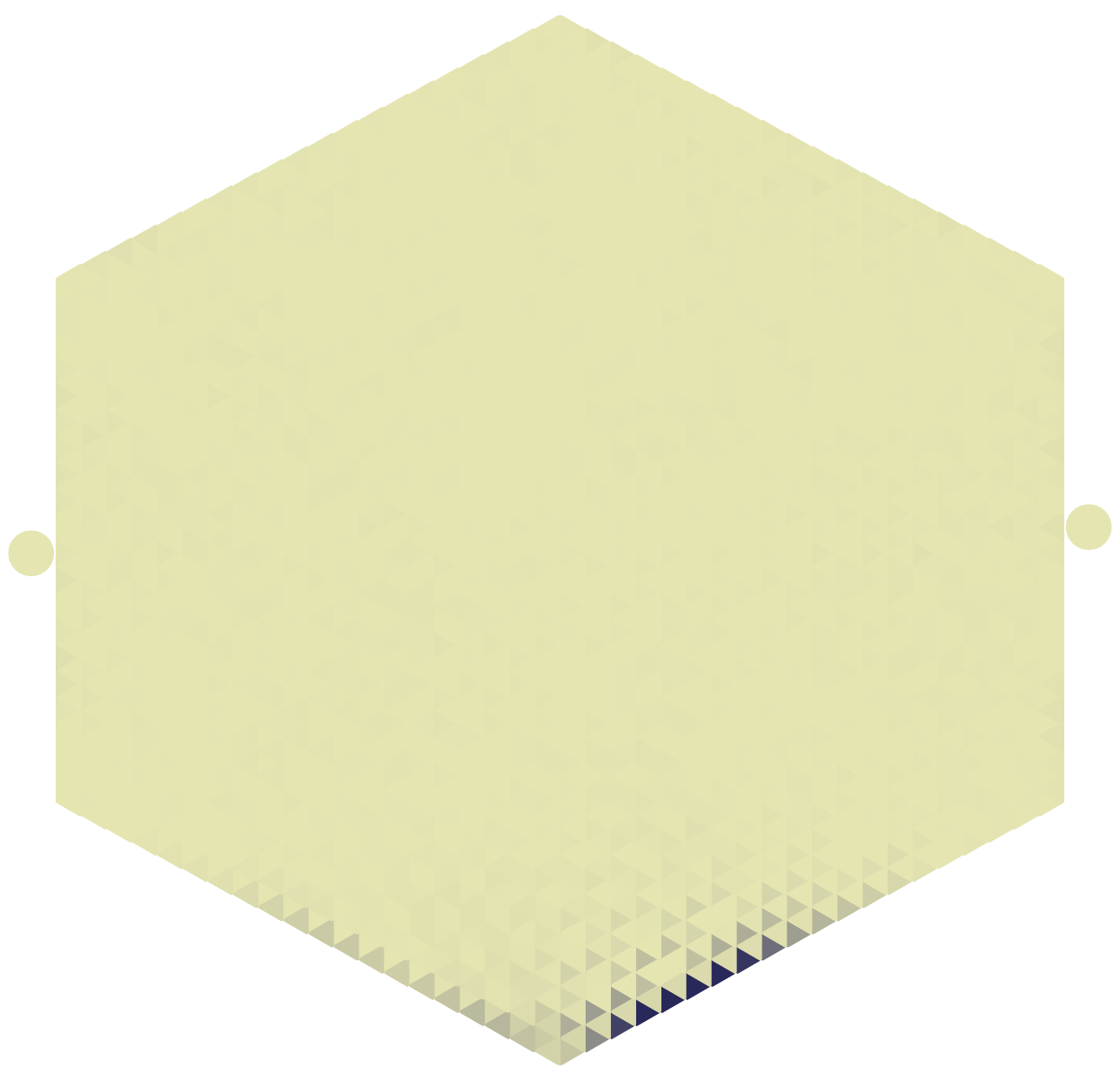}
&\includegraphics[width=0.3\columnwidth]{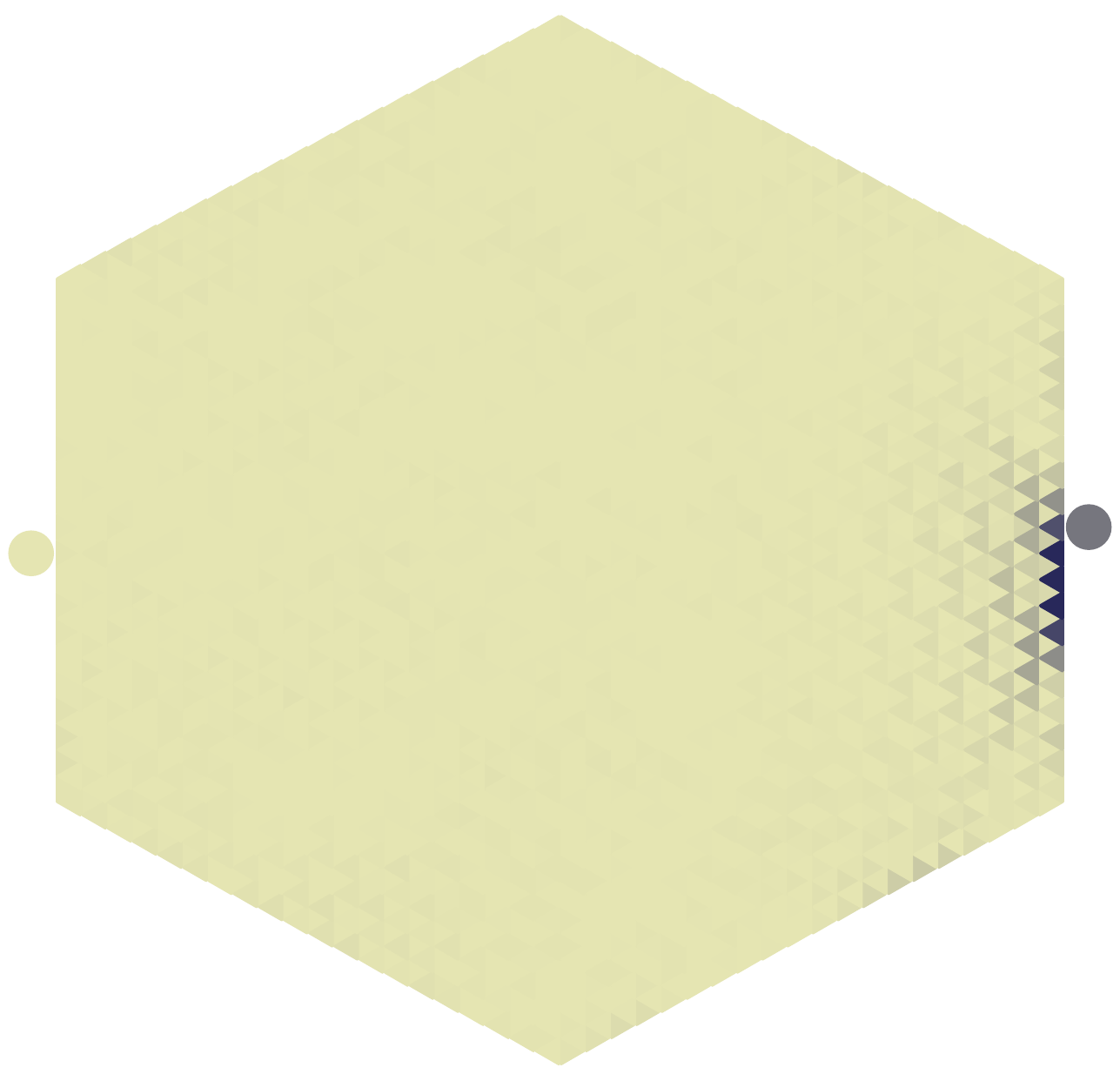}
&\includegraphics[width=0.3\columnwidth]{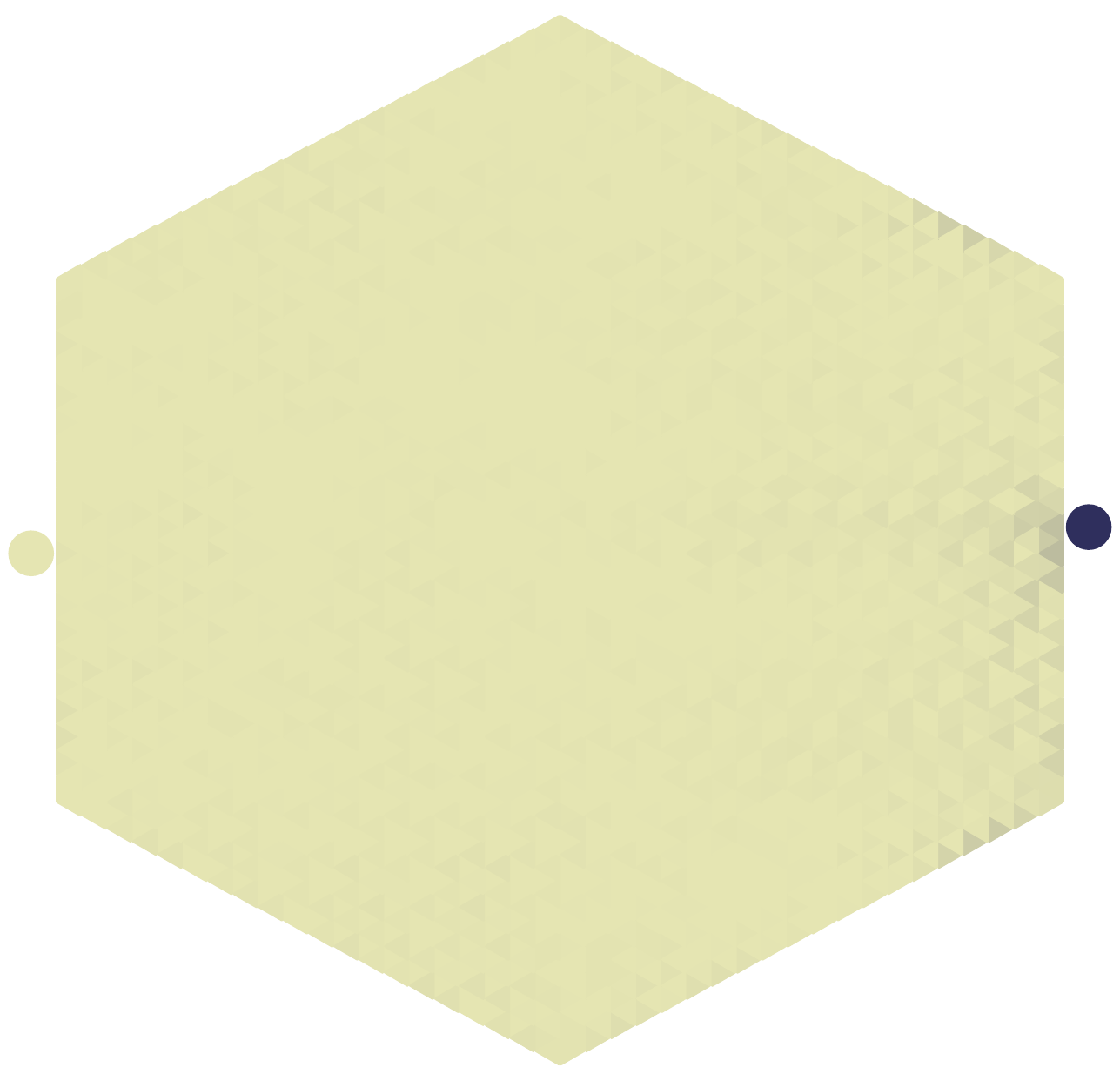}
\end{tabular}
\caption{
Propagation of a Majorana state between two external spins, coupled to the edge of a Kitaev-honeycomb sample: some intermediate positions are shown. Fidelity of 0.964 was reached with each external qubit is coupled to a single edge spin.
The parameters are $\lambda=0.1J$, $v_\textrm{gr}=0.324J$, $\Delta = 0.280J$, the sample contained 2400 sites.
	\label{fig:simul}}
\end{figure}
%%%%%%%%%%%%%%%%%%%%%%%%%%

Further, in a realistic circuit, emulating the Kitaev model, the circuit parameters, such as the spin couplings $J$ and the external fields $h$, may deviate from an ideally uniform distribution. The influence of this kind of disorder, as well as of the noise in these parameters, on the properties of the edge modes and the fidelity of the quantum operations, needs to be studied. Our simulations indicate that for disorder levels consistent with modern-day circuits of superconducting qubits~\cite{GoogleSupremacy}, this influence is not dramatic and does not prevent one from implementation of the described operations~\cite{TTM}. In particular, at sufficiently weak disorder, consistent with the present-day possibilities for superconducting-qubit networks, no fluxes are generated and the edge states remain stable (which is related to their chirality).

%%%%%%%%%%%%%%%%%%%%%%%%%%
\begin{figure}
a)\includegraphics[height=0.23\columnwidth]{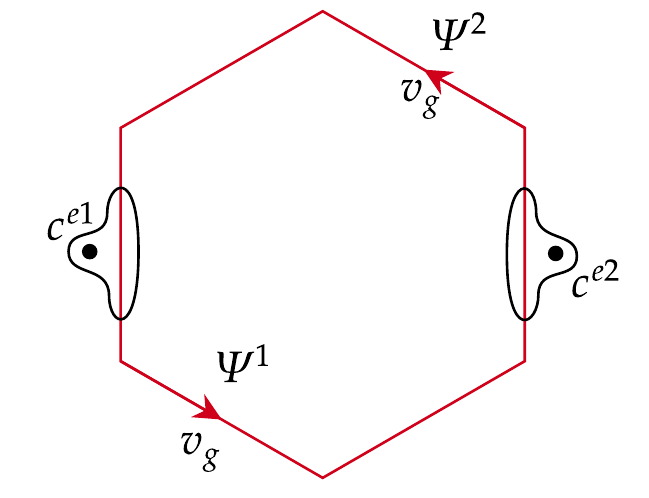}
b)\raisebox{0.06\columnwidth}{\includegraphics[height=0.12\columnwidth]{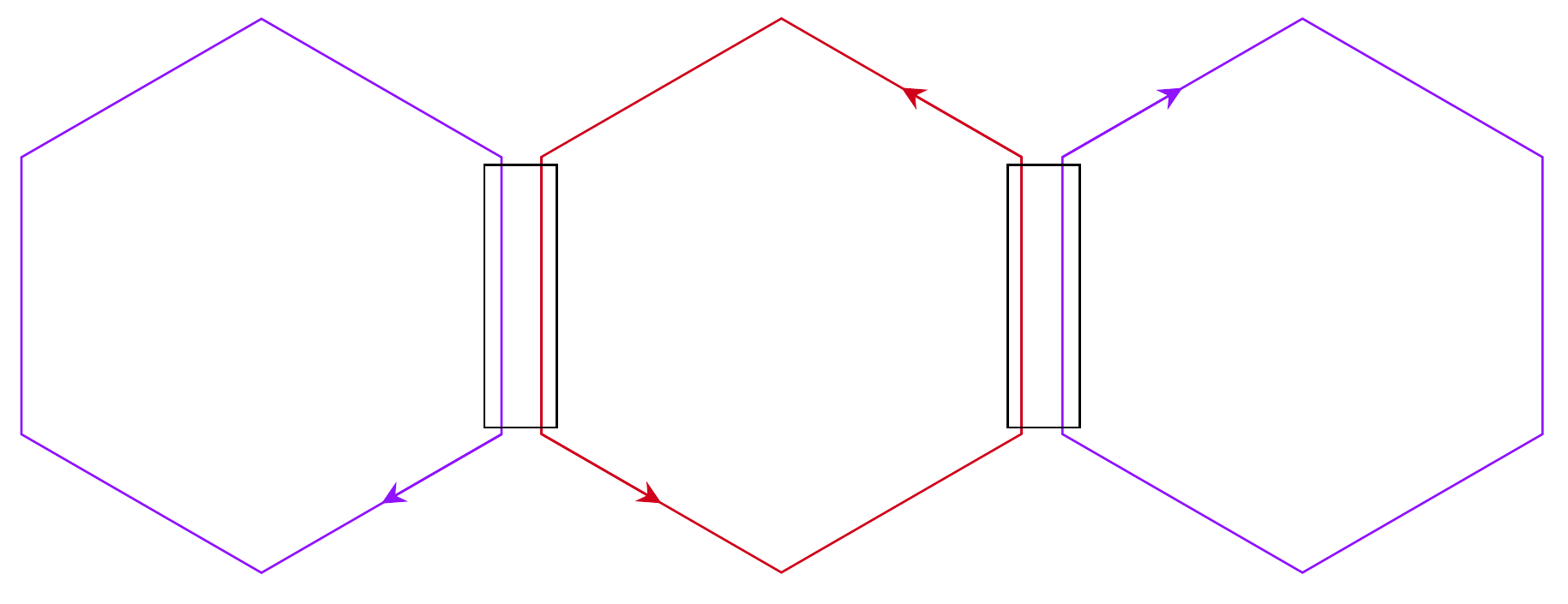}}
c)\includegraphics[height=0.23\columnwidth]{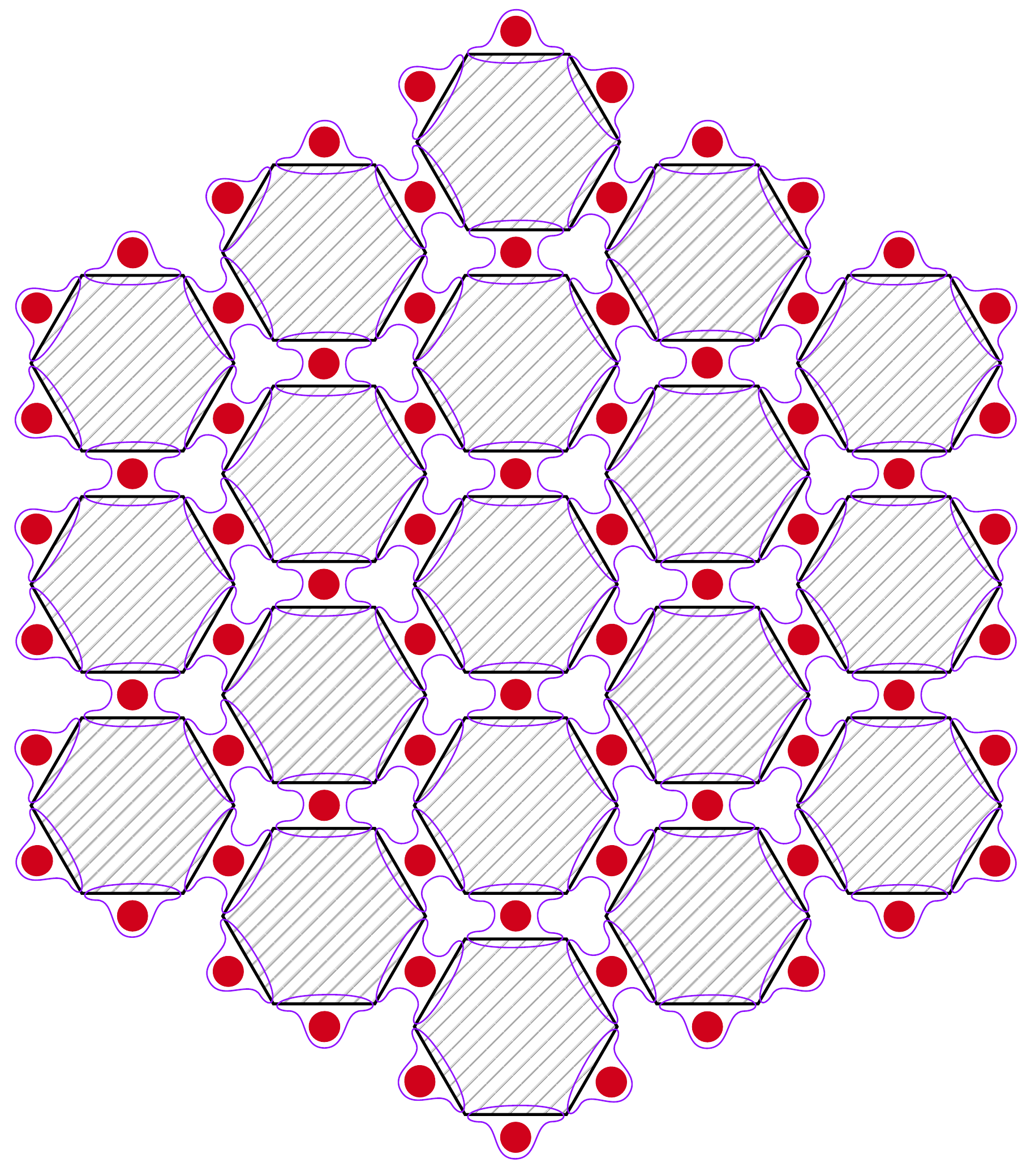}
	\caption{a) Entanglement of two qubits via the edge modes.
	A hexagonal sample with only zigzag edges is shown.
	b) Three samples of Kitaev honeycomb model with different Chern numbers (-1,+1,-1). This design allows for braiding of the `red' Majorana modes along the edge of the central hexagon and, as a result, for a quantum logic gate.
	c) Many-qubit design, using edge states for quantum-information processing, see text.}
	\label{fig:Schemes}
\end{figure}
%%%%%%%%%%%%%%%%%%%%%%%%%%

Furthermore, we discussed the dynamics of the system within the lowest-energy sector without fluxes. Let us mention that our numerical simulations showed almost no sensitivity of the edge transport to fluxes in the bulk, away from the edge, for the zigzag direction and at vanishing field at the edge. At the same time, fluxes near the edge modify the velocity of the edge mode (quantified via the travel time around the sample perimeter). On one hand, these observations may be viewed as stability of the present scheme towards fluxes, while on the other hand, edge transport can be used as a tool to probe if near-edge vortices are present in the system, cf. flux detection discussed in Ref.~\onlinecite{KlockePRL21}. 

One approach to realization of the circuits discussed above may use superconducting quantum bits, actively studied experimentally. For instance, transmons~\cite{TransmonKoch07} or charge qubits can be coupled in various ways to implement $xx$, $yy$, $zz$ couplings. Certain approaches were suggested~\cite{NoriKitaevModel,Sameti2019}, however, some modifications can be useful in order to achieve the needed parameter regime. Indeed, we have discussed the situation, where all couplings $J_x$, $J_y$, $J_z$ are of the same sign. This can be achieved by coupling all neighboring qubits either directly (capacitively for the $zz$ coupling or inductively for $xx$ and $yy$ couplings) or via an intermediate high-frequency resonator~\cite{Nature99} or qubit~\cite{Averin03} mode. This kind of indirect coupling has been discussed in the literature~\cite{Nature99,Averin03,Hutter06,OliverCoupling18,Blais2021} and used in recent experiments with the Google Sycamore processor~\cite{GoogleCoupling,GoogleSupremacy}. Note that one can also combine direct and indirect couplings for better tunability~\cite{OliverCoupling18,GoogleSupremacy}. It further allows one to control the coupling strength to achieve the needed parameters in the bulk but also during the pulsed coupling to external degrees of freedom at the edges. At the same time, while sufficiently uniform couplings (and local fields) may be achieved in a many-qubit circuit~\cite{GoogleSupremacy}, spatial fluctuations may degrade the fidelity of operations, see below.

Alternatively, instead of artificial systems, one may probe the physics discussed above in `natural' Kitaev materials~\cite{Trebst,Tanaka2022}. While in this case one has a much lower degree of control over individual sites, the system may be much more uniform, which is relevant for the observation of the effects discussed~\cite{TTM}. However, in such natural Kitaev materials one may expect an admixture of other, non-Kitaev spin couplings in the system. The effect of spin coupling satisfying the symmetry of the Kitaev model was analyzed in Ref.~\onlinecite{Balents2016}. These extra interaction terms beyond the Kitaev coupling modify the low-energy fermionic description. Under external magnetic field and translated to the situation of interest here, similar to the coupling in Eq.~(\ref{eq:kappa}), this reduces to the next-nearest neighbor couplings, however, with anisotropic coupling constants $\kappa_{x,y,z}$ depending on the direction.
The strength of these $\kappa$-terms, in addition to $h^3$ terms in the pure Kitaev limit, has contributions linear in the field, $\kappa_\alpha\sim h_\alpha (J_e/J)^2$, where $J_e$ is the strength of the non-Kitaev couplings~\cite{Balents2016}.
These effects have minor quantitative consequences, in particular, the gap in the bulk is replaced by $\Delta = 2\sqrt{3} |\kappa_x+\kappa_y+\kappa_z|$, and this modifies slightly the spectrum and shape of the edge modes.
For instance, Eqs.~(\ref{eq:SpSingleMode}) and~(\ref{eq:epszigzagh=0}) are replaced by
\begin{eqnarray}\label{eq:SpSingleModeInteractions}
\varepsilon &=& -\frac{h_z^2}{J^2}
\frac{\kappa_z\sin q_x + \frac12(\kappa_x+\kappa_y) \tan\frac{q_x}{2}}%
{\cos^2\frac{q_x}{2} -\frac{1}{4} +\frac{h_z^2}{4J^2}}
\,.\\
\varepsilon &=& -4(\kappa_x+\kappa_y+\kappa_z)\sin q_x \,,
\end{eqnarray}
However, this does not change qualitatively neither the properties of the edge modes, nor our conclusions above.

An important issue, especially for the artificial, qubit-based Kitaev lattice, is the influence of disorder in the system.
Indeed, in qubit networks effective local fields and couplings in Eq.~\eqref{eq:fullHam} are typically controlled individually and depend on fabrication details. While sufficiently narrow distribution of circuit parameters can be established~\cite{GoogleSupremacy}, it is important to evaluate stability of the system and our findings to static disorder and time-dependent noise. Analysis in Ref.~\onlinecite{TTM} demonstrates stability of the ground state and sufficiently high fidelity of the quantum gates, described above. 

In particular, stationary disorder in the couplings and local fields may modify properties of the edge modes and even the structure of the quantum state. This has been analyzed, together, with the effect of the non-stationary noise, with the conclusion~\cite{TTM} that at not too strong fluctuations their effect is only weak, and an order-10\% spread in the nominally identical circuit parameters, well within reach of the current technology~\cite{GoogleSupremacy}, should not prevent one from implementation of the quantum-state transmission along Majorana edges in samples of a few hundred qubits. This effect was quantified~\cite{TTM} via fidelity of the qubit gates, performed via the edge as discussed in this article.

We are grateful to A.~Shnirman, K.~Tikhonov, and A.~Wallraff for useful discussions.
This work has been supported by RFBR under No.~20-52-12034 and by the Basic research program of HSE.

%%%%%%%%%%%%%%%%%
\appendix
\section{Spin SWAP operation}
\label{app:SWAP}

We described in Section~\ref{sec:Manip} an exchange operation between an external spin and the edge: in the fermionic language, an exchange $c\to -\psi^\ext$ and $\psi^\ext\to c$ of the external Majorana $c$ and the edge Majorana mode $\psi^\ext$. Here we discuss two aspects related to this operation. First, the coupling between $c$ and $\psi^\ext$ is proportional to the link operator $u \equiv ib_z b_z^\ext =\pm1$ and is thus gauge-dependent. This implies that accurate description of the exchange involves $u$:
\begin{equation}
c\to -u\, \psi^\ext \,,\quad \psi^\ext\to u\, c \,.
\end{equation}
We discuss below, how it can be consistent with gauge-invariant qubit operations. To illustrate this, we discuss at the same time a possibility to perform a two-qubit gate on two external spins coupled to the edge at two different locations. This shows an example of the use of Majorana modes to transfer and process quantum information, and it is of interest to implement it in an experiment with any kind of edge states.

Based on this operation, consider an exchange of the `constituent fermions' for two external qubits (or spins) $s_1$ and $s_2$, coupled to the edge, see Fig.~\ref{fig:Scheme}. For convenience, as above, we consider the spins as parts of the Kitaev lattice and decompose them into $c$ and $b$ fermions.

We perform a sequence of operations, which we describe below, and show that it results in the SWAP gate for $s_1$ and $s_2$. This sequence uses constituent operations, indicated in Fig.~\ref{fig:Scheme}: the exchange of $c^i$ and $\psi$ at the location of either spin $s^i$, effected via the local pulsed $zz$-coupling, indicated in green; the exchange of $c^i$ and $b^i_x$ via a local pulse of $h^i_x$ field, in blue; and similar exchange of $c^i$ and $b^i_y$ via a local pulse of $h^i_y$ field, in blue. 

Indeed, first we turn on the interaction between qubit $s_1$ and the edge to write its $c$-fermion, $c^1$, onto the edge, while the local edge state $\psi^{\ext1}$ would be transferred to $s_1$. After the interaction is turned off, the $c^1$ Majorana fermion would start its travel along the edge and would reach the other spin $s_2$ after some time, see Fig.~\ref{fig:Scheme}. After another exchange operation, now at the location of $s_2$, the $c^1$ would be exchanged with $c^2$. The latter would move along the edge down to $s_1$, and after another exchange there, would be written onto the former location of $c^1$ in this first qubit $s_1$.

Following this procedure, one observes that it produces an exchange of the $c$-operators, and at the same time the edge operator $\psi$ would finally acquire its original position at the edge.
However, both $c^1$ and $c^2$ acquire an extra factor of $u^1 u^2$ due to two local exchanges at positions of $s_1$ and $s_2$ each. Here, $u^1=ib^1_z b^{\ext1}_z$ and $u^2=ib^2_z b^{\ext2}_z$ are the link operators on the links between the two qubits and the edge, cf.~Section~\ref{sec:edge} and Ref.~\onlinecite{KitaevHoneycomb}. Thus, the resulting operation is
\begin{equation}\label{eq:c1c2exch}
c^1\mapsto  u^1 u^2 c^2 \,,\qquad
c^2\mapsto -u^1 u^2 c^1 \,.
\end{equation}
The factor $u^1 u^2$ in this expression appears to imply entanglement between the external spins and the edge, undesirable during a quantum operation. Furthermore, so far only the $c$-part of the first qubit $s_1$ is exchanged with its counterpart at the second qubit $s_2$ (although $b_z$-operators at the qubits also became entangled with the edge in the process).

However, a combination of three such operations, with intermediate application of local magnetic fields to the spins, removes these problems and ensures full exchange of the spin states. Specifically, after the operation (\ref{eq:c1c2exch}), one can apply local $h_x$ fields (indicated by blue color in Fig.~\ref{fig:Scheme}) at both qubits to exchange $c^1$ with $b^1_x$ and $c^2$ with $b_x^2$, respectively. Then, another operation (\ref{eq:c1c2exch}) is applied. Finally, after subsequent local exchanges of $c^1$ with $b^1_y$ and $c^2$ with $b_y^2$ via local $h_y$ fields (indicated by blue color in Fig.~\ref{fig:Scheme}), another (\ref{eq:c1c2exch}) is effected. As a result of all these manipulations, the external Majoranas $c$, $b_x$, $b_y$ at the external spin $s_1$ are exchanged with their counterparts at $s_2$, with multiplication by $\pm u^1 u^2$ (in analogy with Eq.~(\ref{eq:c1c2exch}) for $c^i$'s).

%%%%%%%%%%%%%%%%%%%%%%%%%%
\begin{figure}
\includegraphics[width=0.8\columnwidth]{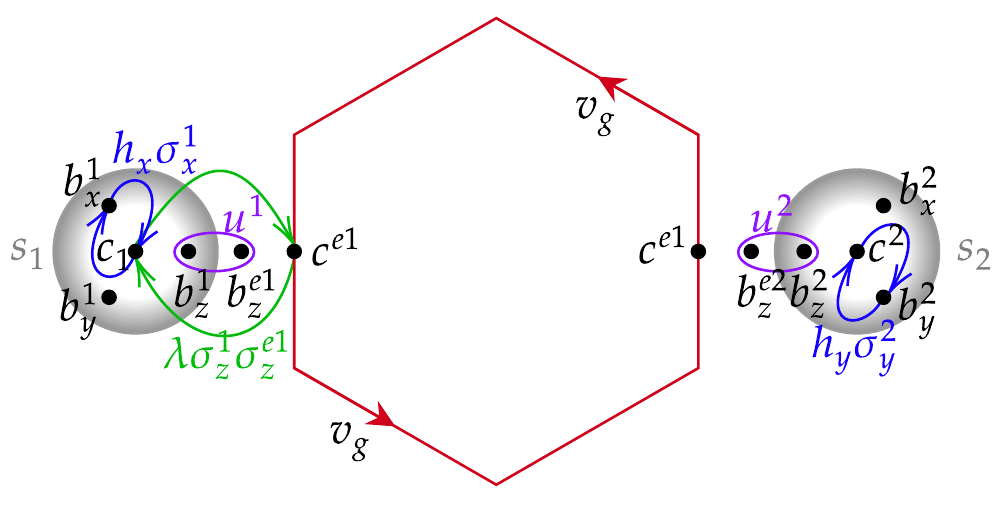}
\caption{Two external spins $s_1$ and $s_2$ are coupled to the edge. The quantum states of the qubits $s_1$ and $s_2$ can be exchanged with an operation constructed from local Majorana exchanges, see text: A pulse of local $zz$-coupling induces an exchange of $c$-fermions, shown in green. An exchange of $c^1$ and $b^1_x$ is effected by a local $h_x^1$ field, shown in blue. Similarly, local $c^i$-$b^i_{x,y}$ exchanges are produced by pulsed local fields $h_{x,y}^i$.}
\label{fig:Scheme}
\end{figure}
%%%%%%%%%%%%%%%%%%%%%%%%%%

While we did not follow the evolution of $b_z^i$'s, it can be deduced from the physical constraints $D^i=+1$. In other words, one can see that as a result of these manipulations, the physical operators evolve according to
\begin{eqnarray}
\sigma^1_x &\leftrightarrow& \sigma^2_x \,,\\
\sigma^1_y &\leftrightarrow& \sigma^2_y \,,\\
D^1\sigma^1_z &\leftrightarrow& D^2\sigma^2_z \,,
\end{eqnarray}
where $D^i=b^i_x b^i_y b^i_z c^i$, cf.~Section~\ref{sec:edge} and Ref.~\onlinecite{KitaevHoneycomb}. Since, $D^{\ext1}$ and $D^{\ext2}$ commute with the Hamiltonian, they conserve their values. For physical states, $D^{\ext1}=D^{\ext2}=1$, hence, in the subspace of physical states the described operations induce the exchange of spins,
\begin{equation}
\sigma^1_\alpha \leftrightarrow \sigma^2_\alpha \,,
\end{equation}
importantly, without any gauge dependence.
In the qubit language, the quantum states of the qubits are swapped, which is known as quantum teleportation. Note also that, importantly, the edge mode disentangles from the external spins after the complete operation.

\bibliographystyle{apsrev4-2}
\bibliography{1dkm}

\end{document}